\documentclass[%
 reprint,
superscriptaddress,
 amsmath,amssymb,
 aps,
 prl,
]{revtex4-1}

\usepackage{graphicx}
\usepackage{dcolumn}
\usepackage{bm}
\usepackage{natbib}
\usepackage{hyperref}
\usepackage{siunitx}
\usepackage[caption=false]{subfig}
\usepackage{booktabs}
\usepackage[export]{adjustbox}
\usepackage{multirow}
\usepackage{physics}
\usepackage{xcolor}
\usepackage{accents}

\newcommand{\zm}[1]{#1}

\begin{document}

\title{Fast mixed-species quantum logic gates for trapped-ion quantum networks}

\author{Zain Mehdi}
 \email{zain.mehdi@anu.edu.au}%
 \affiliation{Department of Quantum Science, Research School of Physics, Australian National University}%
  \author{Varun D. Vaidya}
 \affiliation{IonQ, Inc., College Park, MD, USA}%
\author{Isabelle Savill-Brown}
 \affiliation{Department of Quantum Science, Research School of Physics, Australian National University}%
 \author{Phoebe Grosser}
 \affiliation{Department of Quantum Science, Research School of Physics, Australian National University}%
   \author{Alexander K. Ratcliffe}
 \affiliation{IonQ, Inc., College Park, MD, USA}%
  \author{Haonan Liu}
 \affiliation{IonQ, Inc., College Park, MD, USA}%
  \author{Simon A. Haine}
 \affiliation{Department of Quantum Science, Research School of Physics, Australian National University}%
\author{Joseph J. Hope}
 \affiliation{Department of Quantum Science, Research School of Physics, Australian National University}%
  \author{C. Ricardo Viteri}
 \affiliation{IonQ, Inc., College Park, MD, USA}%

\date{\today}

\begin{abstract}
Quantum logic operations between physically distinct qubits is an essential aspect of large-scale quantum information processing. We propose an approach to high-speed mixed-species entangling operations in trapped-ion quantum computers, based on mechanical excitation of spin-dependent ion motion by ultrafast pulsed lasers. We develop the theory and machine-design of pulse sequences that realise MHz-speed `fast gates' between a range of mixed-isotope and mixed-species ion pairings with experimentally-realistic laser controls. We demonstrate the robustness of the gate mechanism against expected experimental errors, and identify errors in ultrafast single-qubit control as the primary technical limitation. \zm{The proposed mixed-species gate mechanism can be used for fast transfer of quantum information between specialized qubits and quantum memories, which we show enables the protection of matter-photon interfaces against rapid spin dephasing in optical networks of trapped-ion processors. }
\end{abstract}


\maketitle


\emph{Introduction.---} Scalable quantum information processing (QIP) requires the integration of distinct physical qubits for specialized applications. In trapped ion systems, quantum logic operations between distinct atomic species or isotopes underpin a wide range of applications~\cite{Home2013,Tan2015,Bruzewicz2019}: mid-circuit measurement and feedback for quantum error correction~\cite{Negnevitsky2018}, shuttling-based quantum gate teleportation~\cite{Wan2019}, low-excitation transport of qubits between segmented trapping zones~\cite{Tan2015,Lancellotti2023}, quantum logic spectroscopy~\cite{Tan2015}, and photonic interconnects~\cite{Monroe2014}.

Mixed-species entangling gates have previously been demonstrated using optical fields to provide qubit-state-dependent forcing of spectroscopically-resolved vibrational modes of the ion crystal~\cite{Tan2015,Inlek2017,Negnevitsky2018,Bruzewicz2019,Stephenson2020,Hughes2020,Drmota2023,Main2025}. While these approaches are reaching fidelities comparable to state-of-the-art same-species gates (Ref.~\cite{Hughes2020} achieved a two-qubit (2Q) gate fidelity of $99.8\%$), the underlying gate mechanisms rely on the Lamb-Dicke regime which precludes high-fidelity 2Q gates with operation times significantly faster than the motion of the ions~\cite{Schafer2018FastQubits}. This limits typical entangling rates to $\mathcal{O}(10)$kHz~\cite{Ballance2016}, with longer operation times required for error-robust schemes~\cite{Bentley2020,Leung2018}. Furthermore, these gate schemes become harder to implement for long-ion chains with crowded motional spectra~\cite{Leung2018,Wu2018,Figgatt2019,Bentley2020}, as well as mixed-species crystals with significant mass imbalances ($\gtrsim 10\%$)~\cite{Sosnova2021}, which constrains the design of hybrid trapped-ion QIP architectures.

The speed limitation of current mixed-species gate protocols limits high-fidelity transfer of quantum information between specialized qubits which are more sensitive to decoherence than hyperfine qubits typically used for computation. An important example is the transfer of quantum information from a noisy (magnetic-field sensitive) network qubit to a memory (hyperfine) qubit in ion-photon interfaces~\cite{Drmota2023,Drmota2024,Main2025}, which must be orders of magnitude faster than the typical network-qubit dephasing rates of $1-10$kHz~\cite{Tan2015,Inlek2017,Stephenson2020,Drmota2023} to retain high-fidelity quantum computation across an optical network of trapped-ion quantum processors~\cite{Monroe2014,Nigmatullin2016,Bock2018,Stephenson2020,Drmota2023,Drmota2024,Saha2024}. \zm{This technique is essential for distributed quantum computing with trapped ions~\cite{Monroe2014,Main2025}.} Further applications include fast qubit readout using spectrally isolated ancillas, essential for real-time quantum error correction protocols~\cite{Negnevitsky2018}, and rapid entangled state preparation between logic and clock ions for frequency metrology~\cite{Schulte2016}.

In this Letter, we propose an alternative approach to mixed-species quantum logic operations based on mechanical excitation of ion motion using impulsive spin-dependent kicks (SDKs) on nanosecond timescales. We identify high-fidelity gate solutions for a range of experimentally-relevant species pairings, demonstrating that high-fidelity 2Q operations can be performed in hundreds of nanoseconds with tens of impulsive SDKs and experimentally-accessible laser repetition rates. We demonstrate that the mixed-species fast gate protocol is applicable to a broad range of mixed-species pairings relevant to various QIP applications. As an example, we show that fast mixed-species gates enable the high-fidelity transfer of quantum information between between short-lived Zeeman qubits and long-lived hyperfine qubits, which paves the path for high-speed and high-fidelity quantum networks of trapped-ion qubits connected by photonic links.

\emph{Theory of mixed-species fast gates.---} Here we develop the theory of fast entangling gates between mixed-species ion pairs, based on SDKs delivered to each ion by counter-propagating $\pi$-pulse pairs. This mechanism has been extensively studied for same-species crystals~\cite{Garcia-Ripoll2003,Duan2004,Garcia-Ripoll2005,Bentley2013,Taylor2017,Ratcliffe2018,Gale2020,Ratcliffe2020,Mehdi2021f,Mehdi2020b:2D,Torrontegui_2020,Wu2020b} and Bell-state preparation has been experimentally demonstrated with two $^{171}$Yb$^+$ ions (in $20\mu$s and $76\%$ fidelity)~\cite{Wong-Campos2017a}. The mixed-species protocol is distinguished from same-species approaches by allowing for SDKs to be implemented on different transitions for each of the two ions: $\hat{U}_{\rm SDK}(k^{(1)},k^{(2)}) = \prod_{j=1,2} \hat{U}_{\pi}^{(j)}(k^{(j)}) \hat{U}^{(j)}_\pi(-k^{(j)}) $, where $\hat{U}_\pi^{(j)}(k) = \hat{\sigma}_x^{(j)}e^{i k^{(j)} \hat{x}^{(j)}\hat{\sigma}_z^{(j)}}$ is the unitary for an ideal $\pi$-pulse delivered to the $j$-th ion with transition wavelength $\lambda^{(j)} = (2\pi)/k^{(j)}$. 

The fast gate mechanism consists of impulsive SDKs (taken to be effectively instantaneous with respect to the ion motion~\zm{\cite{supp}}), interspersed by free evolution of the ions~\cite{supp}. We will allow for the direction of the pulses to be switched, described by inverting the sign of the wavevectors, i.e. $\hat{U}_{\rm SDK}(-k^{(1)},-k^{(2)})$, such that SDKs arriving at times $\{t_m\}$ are described by wavevectors $z_m k^{(j)}$ where $z_m = \pm 1$ describes the direction of the $m$-th kick. The total unitary for the gate evolution is then $\hat{U}_{\rm G} = \prod_m  \hat{U}(t_{m+1}-t_m)\hat{U}_{\rm SDK}(z_mk^{(1)},z_mk^{(2)})$. Up to a global phase, this is equivalent to the expression~\cite{supp}
\begin{align}
 \label{eq:GateUnitary}
 	\hat{U}_{\rm G} &= e^{i \Theta \sigma_z^{(1)}\sigma_z^{(2)}}\prod_{\alpha=1}^L\hat{D}_\alpha(\beta_\alpha^{(1)}\hat{\sigma}_z^{(1)}+\beta_\alpha^{(2)}\hat{\sigma}_z^{(2)}) \,,
 \end{align}
 where $\beta_\alpha^{(j)} = 2i \eta_\alpha^{(j)}b_\alpha^{(j)}\sum_m z_m e^{i\omega_\alpha t_m}$ characterises the final motional state of each mode ($\hat{D}_\alpha(\beta)$ is the displacement operator on mode $\alpha$ with amplitude $\beta$). Here $\vec{b}_\alpha$ and $\omega_\alpha$ are respective eigenvector and secular eigenfrequency of the $\alpha$-th motional mode, and $\eta_\alpha^{(j)} \equiv k^{(j)}\sqrt{\hbar/(2m^{(j)}\omega_\alpha)}$ is the corresponding ion-dependent Lamb-Dicke parameter with $m^{(j)}$ as the $j$-th ion's mass. The two-qubit phase $\Theta$ in Eq.~\eqref{eq:GateUnitary} is given by~\cite{supp}:
 \begin{align}
 	\label{eq:EntanglingPhase}
 	\Theta = 8\sum_{\alpha} b_\alpha^{(1)}b_\alpha^{(2)}\eta_\alpha^{(1)}\eta_\alpha^{(2)}\sum_{m=2}^\mathcal{N}\sum_{n=1}^{m-1}z_nz_m\sin(\omega_\alpha\delta t_{mn})
 \end{align}
for $\delta t_{mn} \equiv t_m -t_n$ and $\mathcal{N}$ total SDKs. Provided the motional states are restored at the end of the gate operation (i.e. $\beta_\alpha^{(j)} =0$), Eq.~\eqref{eq:GateUnitary} realises a $\hat{\sigma}_z\otimes \hat{\sigma}_z$ phase gate, which is maximally entangling when $\Theta = \pi/4$. We note that the derivation of Eqs.~\eqref{eq:GateUnitary} and \eqref{eq:EntanglingPhase} does not make use of a weak-coupling approximation (see Supplementary Material~\cite{supp}), such that the gate mechanism is performant outside the Lamb-Dicke regime (distinctive from spectroscopic gate mechanisms~\cite{Ballance2016}). 

Imperfect phase accumulation ($\Theta\neq \pi/4$) and imperfect motional restoration ($\beta_\alpha^{(j)} \neq 0$) contributes to a non-zero 2Q gate error that can be quantified by the state-averaged infidelity~\cite{supp}:
\begin{align}
\label{eq:Infidelity_Truncated}
 \frac{3\; \varepsilon_{\rm av}}{2} = |\Delta \Phi|^2+\sum_\alpha \left(\bar{n}_\alpha+\frac{1}{2}\right)\left(|\beta_\alpha^{+}|^2 + |\beta_\alpha^{-}|^2\right )\,,
\end{align}
 where $\Delta\Phi = \Theta- \pi/4$, $\bar{n}_\alpha$ is the mean thermal occupancy of each mode prior to the gate, and $\beta_\alpha^\pm = \beta_\alpha^{(2)} \pm \beta_\alpha^{(1)}$. Note that for perfect motional restoration, $\beta_\alpha^\pm =0$, the fast gate mechanism is completely insensitive to the temperature of the system. Unless otherwise stated, we assume a constant mode occupation of $\bar{n}_\alpha = 1$ for all calculations presented in this work, \zm{though Eq.~\eqref{eq:Infidelity_Truncated} demonstrates the results hold for much larger mode occupations (provided $|\beta^{\pm}_\alpha|$ is sufficiently small)}.

\zm{We identify high-fidelity fast gate schemes by numerical minimization of Eq.~\eqref{eq:Infidelity_Truncated} with respect to SDK timings ($\{t_k\}$) and directions ($\{ z_k=\pm 1\}$) using a two-stage global algorithm~\cite{Gale2020} summarized in Appendix A. To enable experimental implementation with existing laser controls, we impose a $200$~MHz bandwidth limit on the SDK sequence (i.e. a minimum separation of $5$ns between consecutive SDKs) and further add penalties to minimize the total number of SDKs in the gate $\mathcal{N}$. These modifications to previous theoretical fast gate design~\cite{Gale2020,Mehdi2020b:2D,Mehdi2021} allows us to find high-fidelity sub-microsecond gate solutions with tens of SDKs that can be implemented directly from a mode-locked pulsed laser without the overhead of engineering precise optical delays between pulses~\cite{Bentley2013,Mizrahi2013b,Wong-Campos2017a}. Further details are given in Appendix A.}

\begin{figure*}[t!]  \includegraphics[width=2\columnwidth]{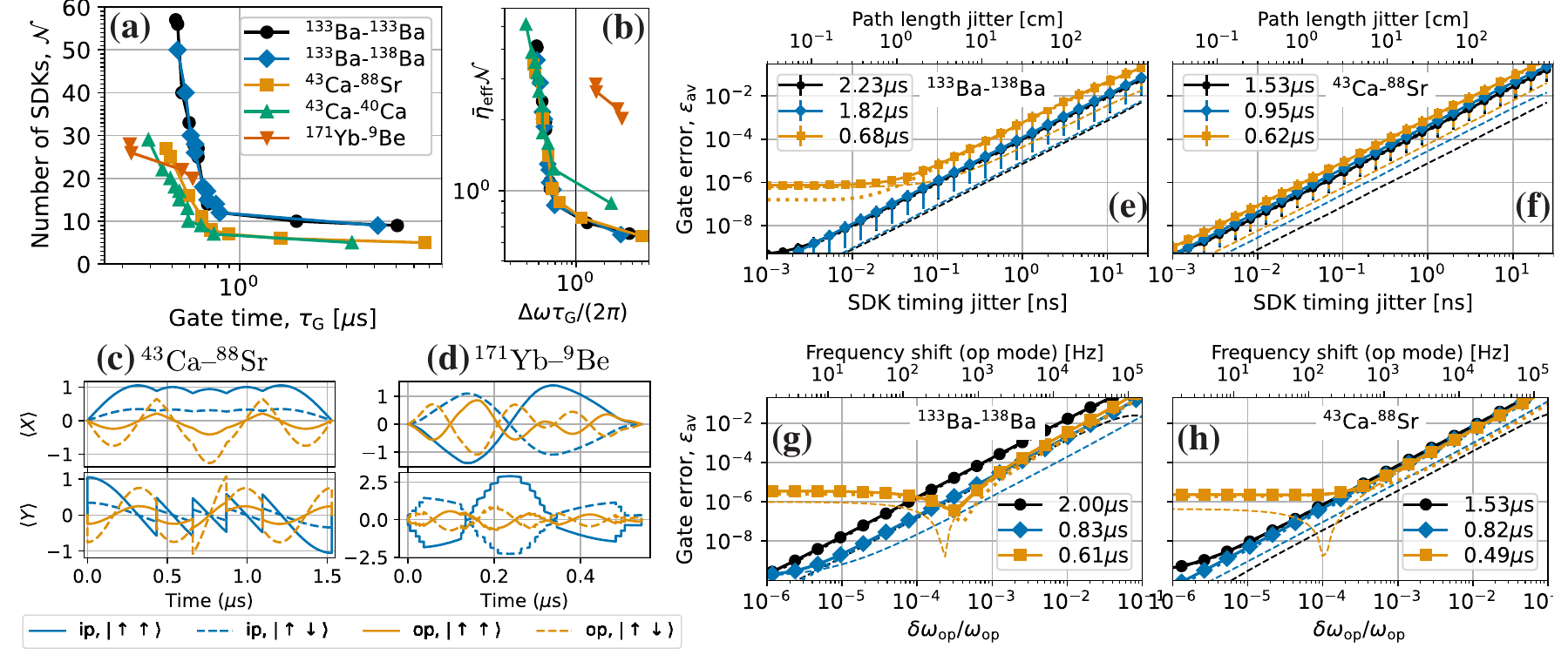}
\caption{\label{fig:TwoIons_Longitudinal}  (a) Fast gate solutions with state-averaged fidelities above $99.9\%$, assuming idealised SDKs, demonstrating a trade-off between the gate time ($\tau_{\rm G}$) and number of SDKs ($\mathcal{N}$). (b) A universal trend is revealed under the rescaling $\mathcal{N}\rightarrow \bar{\eta}_{\rm eff}\mathcal{N}, \tau_{\rm G}\rightarrow \Delta\omega \tau_{\rm G}/(2\pi)$, with the exception of the large-mass-imbalance pair $^{171}$Yb--$^{9}$Be. \zm{(c-d) The motional dynamics of exemplary gate solutions in terms of the means of the mode quadratures, $\hat{X}=(\hat{a}+\hat{a}^\dag)/\sqrt{2}$ and $\hat{Y}=i(\hat{a}^\dag-\hat{a})/\sqrt{2}$, for the in-phase (ip) and out-of-phase (op) motional modes. (c) illustrates a $1.5\mu$s gate between $^{43}$Ca and $^{88}$Sr ($\varepsilon_{\rm av}\approx 2\times 10^{-10}$). (d) shows a $560$ns gate between $^{171}$Yb and $^{9}$Be using $\mathcal{N}=25$ SDKs ($\varepsilon_{\rm av}\approx 10^{-4}$).} (e-f) Sensitivity of select gate solutions against timing jitter in the SDK pulse sequence, with the gate error averaged over $10^{4}$ noise realisations. Errorbars indicate the standard deviation in the ensemble average. (g-h) Impact of frequency drifts of the high-frequency op mode on the state-averaged gate error, for select gate solutions.}
\end{figure*}

\emph{Results.---} \zm{The theory of mixed-species fast gates applies generally to any ion species which can be manipulated by optical fields to implement high-fidelity SDKs, either by single-photon transitions on optical qubits~\cite{Heinrich2019,Hussain2023} or Raman transitions between hyperfine~\cite{Campbell2010a,Mizrahi2013b,Johnson2015,Johnson2017} or Zeeman qubits~\cite{Putnam2024a}. As illustrative examples, we consider 2Q gates} between a number of different experimental pairings relevant to quantum networking: $^{133}{\rm Ba}-^{138}$Ba~\cite{Auchter2014,OReilly2024}, $^{43}{\rm Ca}-^{88}{\rm Sr}$~\cite{Bruzewicz2019c,Hughes2020,Drmota2023,Drmota2024}, and $^{43}{\rm Ca}-^{40}{\rm Ca}$~\cite{Ballance2015}. Each pairing contains a hyperfine qubit and Zeeman qubit, such that the former can be used as a quantum memory and the latter can be used for photonic interfaces~\cite{Monroe2014,Nigmatullin2016}. We also consider fast gates between the unusual pair $^{171}$Yb--$^9$Be, to demonstrate that the fast gate mechanism is performant for large mass-imbalance pairs, which are highly limiting for spectroscopic gate protocols~\cite{Sosnova2021}.

\begin{figure}[t!]
\centering
\includegraphics[width=\linewidth]{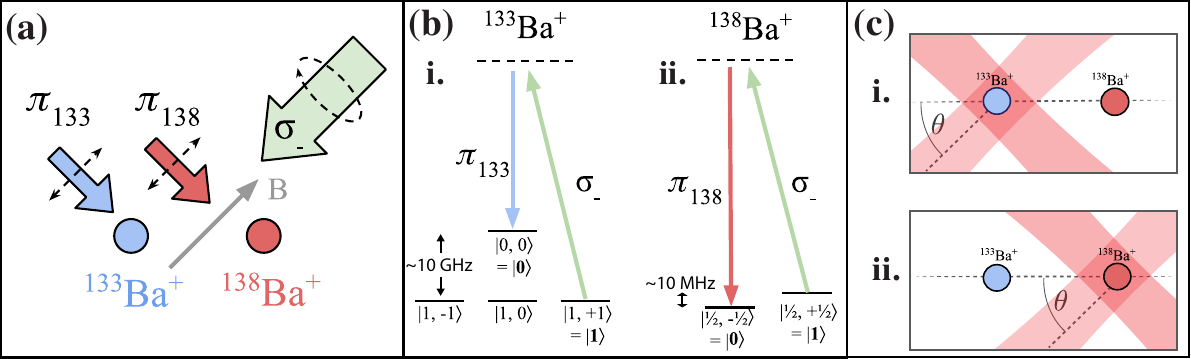}
    \caption{Implementation of SDKs in a dual-species $^{133}$Ba$^+$--$^{138}$Ba$^+$ chain. (a) Configuration of Raman beam orientations, polarizations and magnetic field directions relative to the 2-ion chain. (b) i. Energy levels of the $6s_{1/2}$ manifold in $^{133}$Ba$^+$ illustrating the two qubit states and Raman couplings between them. ii. Energy levels and Raman couplings between the qubit states in the $6s_{1/2}$ manifold of $^{138}$Ba$^+$. (c) SDKs can be implemented asynchronously, e.g. first an SDK is implemented on $^{133}$Ba$^+$ using nanosecond sequences of ultrafast pulses (i.) followed by a single-pulse SDK on $^{138}$Ba$^+$ (ii.), as described in the main text.}
    \label{fig:SDK_Diagram}
\end{figure}

In each case, we assume the hyperfine/Zeeman qubits are stored in the $s_{1/2}$ ground-state manifold, and that SDKs are implemented on the axial modes of the trap using counter-propagating Raman beams tilted at an angle of $\theta=\pi/4$ from the longitudinal axis of the trap~\cite{Putnam2024a} [see Fig.~\ref{fig:SDK_Diagram}(c)] such that each the effective wavevector of the two-photon transition is $k_{\rm eff}^{(j)} = \pm \sqrt{2}k^{(j)}$. We will provide an explicit example of a SDK implementation for the $^{133}{\rm Ba}-^{138}$Ba pairing later in this manuscript. For concreteness, we assume the hyperfine qubit experiences a trapping frequency $\omega_{0,{\rm H}}=2\pi\times 1$MHz; due to the mass-dependence of the RF-pseudopotential, the Zeeman qubit will experience a trapping frequency of $\omega_{0,\rm{Z}} = \omega_{0,\rm{H}}\sqrt{m^{(\rm{H})}/m^{(\rm{Z})}}$. \par

Figure \ref{fig:TwoIons_Longitudinal}(a) presents gate solutions that have state-averaged fidelities above $99.9\%$ before the inclusion of errors in the SDKs \zm{(some solutions have infidelities as low as $10^{-12}$)}, and demonstrates that sub-microsecond gate speeds are achievable with small numbers of SDKs ($\mathcal{N}\sim 5-30$) across all mixed-species pairs considered. \zm{This is a significant improvement over previous theoretical work that reported a significantly larger number of pulses and SDK repetition rates to achieve similar gate times (see Appendix for a comparison with previous theoretical works~\cite{Gale2020,Mehdi2021f}).} This further advances prospects for experimental realisation of fast gates: Ref.~\cite{Wong-Campos2017a} used $10$ SDKs to experimentally demonstrate a $20\mu$s entangling operation. Fig.~\ref{fig:TwoIons_Longitudinal}(a) demonstrates that an order of magnitude increase in gate speed can be achieved using similar numbers of SDKs. The fastest gate solutions are found for the high-mass-imbalance pair, $^{171}$Yb--$^9$Be, with gate operation times as low as $300$ns achievable with $\mathcal{N}\geq 25$ SDKs.

For pairings of $^{43}{\rm Ca}-^{88}{\rm Sr}$ and $^{43}{\rm Ca}-^{40}{\rm Ca}$, Fig.~\ref{fig:TwoIons_Longitudinal}(a) demonstrates that MHz-speed gates are achievable with as few as $5-10$ SDKs. We find a $1.8\mu$s gate solution with only $\mathcal{N}=5$ SDKs for the pair $^{43}{\rm Ca}-^{88}{\rm Sr}$ \zm{(Fig.~\ref{fig:TwoIons_Longitudinal}(c))}, which is promising for near-term experimental demonstrations (see discussion on imperfect SDKs below). A modest increase to $\mathcal{N}\approx 20$ SDKs enables high-fidelity gate solutions with minimal intrinsic error and operation speeds below $500$ns for both of these pairings. This is comparable to the fastest two-qubit gate experimentally demonstrated on trapped-ion systems $480$ns~\cite{Schafer2018FastQubits}, which was subject to large errors ($\sim 40\%$) due to out-of-Lamb-Dicke-regime effects~\cite{Schafer2020b}. 

\zm{Fig.~\ref{fig:TwoIons_Longitudinal}(b)} indicates that distinct ($\tau_{\rm G}$,~$\mathcal{N}$) trends for each mixed-species pair collapse onto a universal tendency under the rescaling $\mathcal{N}\rightarrow \bar{\eta}_{\rm eff}\mathcal{N}, \tau_{\rm G}\rightarrow \Delta\omega \tau_{\rm G}/(2\pi)$, where $\bar{\eta}_{\rm eff}^2 = |b^{(1)}b^{(2)}|\eta^{(1)}\eta^{(2)}$ characterises the effective SDK coupling to the ion pair (\emph{c.f.} Eq.~\eqref{eq:EntanglingPhase}), and $\Delta \omega = \omega_{\rm op}-\omega_{\rm ip}$ is the frequency splitting between the out-of-phase (op) and in-phase (ip) motional modes of the two-ion crystal \zm{(explicit expressions given in Appendix A)}. Specifically, solutions with gate operation times $\tau_{\rm G}\ll 2\pi/\Delta \omega$ collapse onto the same curve. The only ion pair to deviate from this universal trend is $^{171}$Yb--$^9$Be, which has distinct motional trajectories due to the large frequency splitting of $\Delta \omega/\omega_{0,\rm{H}}\approx 5$ (the other pairs considered fall close to the same-species result, $\Delta \omega/\omega_0=\sqrt{3}-1$). This is highlighted in the motional trajectories of an exemplary sub-microsecond gate between $^{171}$Yb--$^9$Be shown in \zm{Fig.~\ref{fig:TwoIons_Longitudinal}(d)}, where there are many oscillations of the high-frequency op mode ($6.21$MHz) during a $560$ns gate operation, while there is only a single oscillation of the low-frequency ip mode ($1.22$MHz) in the same duration.

\emph{Fast gate robustness}.--- \zm{An appealing advantage of fast 2Q gate protocols is a natural insensitivity to external perturbations which are often slow compared to ion motion in typical RF traps. Here we quantify the robustness of the mixed-species fast gate mechanism against expected experimental limitations.}

\zm{\emph{(1) Timing jitter and fast trap noise}. The fast gate mechanism is sensitive to the phase of the free evolution of the motional modes at the arrival time of each SDK, i.e. to fluctuations in $\omega_\alpha t_k$ (\emph{c.f.} Eq.~\eqref{eq:EntanglingPhase}). Noise in this phase can arise due to timing jitter of the SDKs or motional dephasing processes which can be modelled as high-frequency noise ($\gtrsim$MHz) in the motional frequencies (e.g. due to current noise in the trap electrodes). Here we model the former case by adding Gaussian random noise to the timing of each SDK while enforcing a minimum pulse separation of $5$ns (corresponding to a SDK train bandwidth of $200$MHz). Fig.~\ref{fig:TwoIons_Longitudinal}(e-f) demonstrates timing jitters of $\lesssim 1$ns contribute errors of $\mathcal{O}(10^{-4})$ for MHz-speed mixed-species gates for the $^{133}{\rm Ba}-^{138}$Ba and $^{43}{\rm Ca}-^{88}{\rm Sr}$ pairs, with similar results for other ion pairings~\cite{supp}. Given the sub-picosecond timing stability of ultrafast pulsed lasers~\cite{Clark1999}, pulse timing errors are most likely to arrive due to optical path-length variations. As $100$ps timing jitter corresponds to roughly $3$cm path-length, we conclude the fast gate protocols presented here are robust to SDK timing noise. If we consider errors in $\omega_\alpha t_k$ that arise from noise in the trapping frequency instead of pulse timing jitter, we find that trap stabilization below $10^{-3}$ is necessary to ensure that motional dephasing errors remain below $10^{-4}$. In addition, we show in the Supplemental Material~\cite{supp} that the fast gate schemes maintain fidelity despite substantial ($\gtrsim 10\%$) drifts in the SDK repetition rate.}

\emph{(2) Stray fields.} Next we consider the effects of stray electric fields, which can alter the secular frequencies of the RF-pseudopotential and cause mode-dependent shifts in the motional spectra of the system, thereby disturbing the motional dynamics of the gate operation. \zm{Given motional dephasing processes are much slower than the MHz-speed fast gates we consider here (e.g. $4$ms dephasing time measured in Ref.~\cite{Talukdar2016}), we can model the effect of stray fields} as a systematic shift in the mode frequencies. We consider two cases: mode-dependent drift of the op mode (holding the ip mode frequency constant), and frequency drifts common to both the ip and op modes. For the former case, Fig.~\ref{fig:TwoIons_Longitudinal}(g-h) demonstrates that frequency drifts of $\mathcal{O}(1)$kHz induce errors of roughly $10^{-4}$ for a range of MHz-speed gates, with similar results observed for frequency drifts common to both modes~\cite{supp}. Thus, high-fidelity fast gates require that the trap frequencies be stabilized at the $0.1\%$ level, which is easily achievable in existing microfabricated traps~\cite{CharlesDoret2012}. \\

\zm{\emph{(3) Motional heating.} Motional heating of axial modes of ion crystals in RF traps (e.g. due to electric field noise) is a significant source of error in current experimental demonstrations of mixed-species 2Q gates~\cite{Tan2015,Inlek2017,Hughes2020,Drmota2023}. Fast gates are inherently robust to motional heating as their durations are orders of magnitude faster than typical heating rates~\cite{Taylor2017}. Concretely, for a heating rate of $100$~phonons/sec (typical of room-temperature surface-electrode traps~\cite{Harty2014,Bruzewicz2015}) we estimate the induced 2Q gate error ($\epsilon \sim \dot{\bar{n}}\tau_{\rm G}$) to be of the order $10^{-4}$. This provides a pathway to speeding up remote entanglement generation in trapped-ion networks, which currently require significant cooling overheads to mitigate heating due to photon recoil~\cite{Inlek2017,Drmota2023,Main2025}.   }

\zm{\emph{(4) SDK errors.} The key challenge to experimentally realizing high-fidelity fast gates is the control of SDK errors~\cite{Bentley2016}, which lead to errant motional trajectories that contribute 2Q errors that compound with $\mathcal{N}$ -- concretely, the 2Q fidelity can be bounded as $\mathcal{F}(\epsilon_{\pi}) \geq (1-\mathcal{N}\epsilon_{\pi})^2$~\cite{Gale2020}, where $\epsilon_{\pi}$ is a characteristic population transfer error between spin-motional states of a single physical qubit, e.g. $\ket{\downarrow,0}\leftrightarrow\ket{\uparrow,i\eta}$. Current experiments with ultrafast pulses achieve SDK fidelities around $1\%$~\cite{Johnson2017,Putnam2024a}, which limits feasible gate solutions to those with $\mathcal{N}\lesssim 10$. For a $1.8\mu$s gate between $^{43}$Ca--$^{88}$Sr with $\mathcal{N}=5$ SDKs, a two-qubit gate fidelity of $\approx 93\%$ is feasible with current state-of-the-art SDK errors of $\epsilon_\pi \approx 7\times 10^{-3}$~\cite{Johnson2015}.}

\zm{Achieving 2Q gate fidelities of $99.9\%$ requires experimental characterization and suppression of SDK errors at the $10^{-4}$ level. SDKs can be made robust to technical noise sources (e.g. laser intensity fluctuations) by employing adiabatic dark-state transfer protocols such as STIRAP~\cite{Bergmann2019}, which can be benchmarked in a single ion system. For hyperfine qubits, the most significant source of SDK error is off-resonant diffraction of the atomic wavepacket by the Raman beams into higher-order momentum states ($\ket{\downarrow,0}\rightarrow \ket{\uparrow, \pm 2i\eta},\ket{\downarrow \pm 3i\eta},\dots$) due to the Kapitza-Dirac effect~\cite{Gould1986}. This precludes high-fidelity SDK between hyperfine qubit states in a single pulse, due to the lack of frequency separation between the qubit splitting and pulse bandwidth for experimentally feasible laser powers~\cite{Campbell2010a,Mizrahi2013b,Putnam2024a}. Off-resonant diffraction into unwanted momentum modes is sufficient to account for $\sim1\%$ SDK errors in experimental demonstrations of Raman SDKs~\cite{Mizrahi2013}, though higher fidelities are achievable using engineered optical delays or using longer pulse trains~\cite{Mizrahi2013b} at the cost of increased experimental complexity. 
}%

\emph{SDK implementation for $^{133}$Ba-$^{138}$Ba.---} Next, we outline an exemplary implementation scheme for Raman SDKs in a dual isotopes $^{133}$Ba$^+$--$^{138}$Ba$^+$ chain, utilising beams sourced from a single Raman laser, illustrated in Fig.~\ref{fig:SDK_Diagram}(a). The `network' and `memory' qubits are encoded in the $6s_{1/2}$ ground-state manifold of $^{133}$Ba$^+$ and $^{138}$Ba$^+$, respectively, as shown in Fig.~\ref{fig:SDK_Diagram}(b).

Ultrafast spin-motional entanglement of the $^{133}$Ba$^+$ qubit can be realised using a circularly-polarized $\sigma_-$ beam that illuminates both ions and the linearly-polarized $\pi_{133}$ beam that solely addresses the $^{133}$Ba$^+$ qubit. The large hyperfine splitting of $9.926$~GHz in $^{133}$Ba$^+$ allows the $\sigma_-$ and $\pi_{133}$ beams to resonantly address the $\ket{0, 0} \rightarrow \ket{1, +1}$ transition, while minimizing off-resonant couplings to the $\ket{1,-1}$ and $\ket{1,0}$ states (see Fig.~\ref{fig:SDK_Diagram}(b)) and without driving transitions on the $^{138}$Ba$^+$ qubit, which has no hyperfine structure. High-fidelity SDKs on the $^{133}$Ba$^+$ memory qubit can be realised using nanosecond sequences of $\sim 10$ pulses sourced from a single ultrafast laser with a repetition rate of several GHz, utilizing pulse sequences developed in Refs.~\cite{Mizrahi2013,Mizrahi2013b}.\\

SDKs on the $^{138}$Ba$^+$ network qubit may be performed as demonstrated in \cite{Putnam2024a} using $\mathcal{O}(10)$ps pulses of the $\pi_{138}$ beam instead of $\pi_{133}$, in addition to the common $\sigma_-$ beam. Due to the large frequency separation between the pulse bandwidth ($\sim 10$GHz) and the Zeeman qubit splitting ($\sim 10$MHz), SDKs on $^{138}$Ba$^+$ can be performed with single pulses, and can be retroreflected with a delay to implement the unitary in Eq.~\eqref{eq:GateUnitary}~\cite{Putnam2024a}. Furthermore, as the fast gate protocols presented here require only nanosecond control over SDK timings (Fig.~\ref{fig:TwoIons_Longitudinal}(e)), SDKs on $^{138}$Ba$^+$ can be performed asynchronously to SDKs on the hyperfine qubit without significant degradation of the 2Q gate fidelity [see Fig.~\ref{fig:SDK_Diagram}(c)], easing technical constraints for experimental implementation. Alternatively, each resonant $\pi$-pulse can be broken into a nanosecond train of weaker pulses to reduce laser power requirements, or even replaced with composite pulse sequences that are robust to expected noise sources and systematic errors.

\zm{\emph{Application to photonic interface.}--- An advantage of the fast gate mechanism is the ability to transfer quantum information between stable quantum memories and specialized qubits with short coherence times. One example is the storage of ion-photon entanglement in robust quantum memories, which is central to proposals for scalable QIP in trapped-ion networks~\cite{Monroe2014,Nigmatullin2016,Drmota2024}. The fidelity of this protocol is limited by dephasing of the network qubit (typically due to magnetic field noise) during the coherent transfer to the memory qubit~\cite{Stephenson2020,Saha2024}. While this limitation could be addressed with advanced magnetic shielding and stabilization techniques, fast gates offer an appealing alternative by simply performing the qubit swap orders of magnitude faster than state-of-the-art network dephasing rates of $\mathcal{O}(1)$kHz~\cite{Inlek2017,Stephenson2020,Main2025,Saha2024}.
}

As a concrete example, we consider the implementation of an i\textsc{swap} operation (consisting of two $\hat{\sigma}_z\otimes \hat{\sigma}_z$ gates, and single-qubit rotations) between a $^{138}$Ba$^+$ network qubit and a $^{133}$Ba$^+$ memory qubit initialized in $\ket{\downarrow}$~\cite{Drmota2023,Main2025}, where the former has a coherence time of $4$ms as in the experiment of Ref.~\cite{Inlek2017}. With $\mathcal{N}=14$ SDKs, $^{133}$Ba$^+$--$^{138}$Ba$^+$ entangling gates can be performed in approx.~$770$ns (see Fig.~\ref{fig:TwoIons_Longitudinal}(a)), which enables \textsc{swap} gate times of approx. $1.6\mu$s assuming a $60$ns time budget for ultrafast single-qubit operations~\cite{Campbell2010a,Mizrahi2013,Mizrahi2013b}. The error induced by dephasing of the network qubit can then be estimated as $1-e^{-1.6\mu{\rm s}/(4{\rm ms})}\approx 4\times 10^{-4}$. In comparison, current spectroscopic mixed-species gates have operation times of ${\sim}50\mu$s~\cite{Hughes2020,Drmota2023}, which leads to an expected dephasing error of $\approx 4\%$ (assuming a i\textsc{swap} duration of $100\mu$s). Even assuming spectroscopic gate times approaching the state-of-the-art for high-fidelity same-species pairs, $\approx 15\mu$s~\cite{Saner2023}, the dephasing error would still be $\approx 0.7\%$~\footnote{These estimates are consistent with $\mathcal{O}(1\%)$ dephasing error contributions to remote entanglement fidelities observed in experiment~\cite{Stephenson2020,Main2025}.}. \\ 

\zm{Network speeds in trapped-ion systems are currently limited by the speed of remote entanglement generation (hundreds of Hz~\cite{Stephenson2020,OReilly2024}), however technical improvements in future experiments (e.g. improvement of photon collection using integrated optics) could feasibly enable remote entanglement speeds approaching the MHz regime~\cite{Nickerson2014,Knollmann2024}. Therefore in the long-term, fast gate protocols  will become necessary to prevent local entangling gates from becoming bottlenecks in high-speed quantum networking and distributed quantum computing~\cite{Main2025}.}

\emph{Conclusions.---} We have demonstrated the suitability of fast entangling gates based on impulsive SDKs for high-speed and high-fidelity quantum logic gates in a range of mixed-species pairings. Our results indicate that the fast gate mechanism is largely agnostic to the choice of species pair (beyond the ability to perform high-fidelity SDKs on each species independently), which enables multiple specialized qubits to be employed on each processing node while retaining the ability to perform high-fidelity and high-speed quantum logic operations between any mixed-species pair.

\emph{Acknowledgments.}--- This research was undertaken with the assistance of supercomputing resources and services from the National Computational Infrastructure, which is supported by the Australian Government.

\pagebreak
\appendix
\renewcommand{\thefigure}{A\arabic{figure}} 
\renewcommand{\theequation}{A\arabic{equation}} 
\setcounter{equation}{0}
\setcounter{figure}{0}
\section{Appendix A: Gate search details}

The machine-design of the SDK sequence can be described as the optimisation of Eq.~\eqref{eq:Infidelity_Truncated} over the SDK directions $\{z_m\}$ and timings $\{t_m\}$. We employ a two-stage procedure, following the \zm{Generalized Pulse Group (GPG)} scheme developed in Ref.~\cite{Gale2020}. In the first stage, SDKs are constrained to arrive in $N$ groups with uniformly spaced timings, $t_m = \tau_{\rm G}m/N$, where $\tau_{\rm G}$ is the desired gate time. For computational simplicity, free evolution between SDKs within each group is neglected in the first stage, such that optimisation is performed over the parameters
\begin{align}
    \vec{z} &= \{ z_1, z_2, \dots, z_N\} \,,
\end{align}
where the elements are allowed to take any integer values within a given range. For this work, we use between $N=16$ and $N=20$ pulse groups in the first optimisation stage, in order to provide sufficient timing freedoms for high-fidelity solutions, and  bound $|z_m|\leq 5$ to limit the number of SDKs in each group. We employ a heuristic approach for optimising the integer parameters $z_m$: a large ensemble ($10^{3}-10^{4}$) of randomly sampled initial conditions are propagated by gradient-descent local minimisations, while treating $z_m$ as floating-point numbers. The final solution of each local minimisations are then integerised, and passed onto the second stage of the optimisation.\\
In the second stage of optimisation, the timings of each SDK group are locally optimised to fine tune the motional trajectories of the ions. In this stage, the finite SDK bandwidth is incorporated by imposing a minimum time separation of $5$ns between consecutive pulses in the same group (corresponding to a $200$MHz SDK repetition rate). An additional $\tanh$ penalty is added to the cost function to prevent temporal overlap of pulse groups. The second stage of optimisation is performed on all candidate solutions identified in the first stage: the optimal SDK sequence is identified as the  pulse with the lowest gate error given by Eq.~\eqref{eq:Infidelity_Truncated}.

\subsubsection{Comparison to previous theoretical work}
\zm{The design of SDK sequences to implement fast gates has seen significant theoretical development over the past two decades~\cite{Garcia-Ripoll2003,Garcia-Ripoll2005,Duan2004,Bentley2013,Ratcliffe2018,Gale2020,Mehdi2021f,Torrontegui_2020}. For experimental implementation, especially with hyperfine qubits and Raman beams, the most suitable schemes will minimize the number of SDKs to limit compounding errors~\cite{Garcia-Ripoll2003,Duan2004,Bentley2013,Bentley2016,Wong-Campos2017a,Gale2020,Mehdi2021}. These schemes should also utilize pulse sequences with SDKs separations resolvable within a few nanoseconds, avoiding the need for complex optical delay engineering~\cite{Bentley2013,Wong-Campos2017a}. As compared to the present work, where we identify fast gate protocols that require $\mathcal{N}\sim 10$ SDKs with at least $5$ns delay between consecutive kicks, previous theoretical work reported larger numbers of pulses ($\mathcal{N}\sim 100$) and/or larger SDK repetition rates ($\sim 1$GHz) to find gate solutions of similar (MHz) speeds and fidelities for same-species ion pairs~\cite{Gale2020,Torrontegui_2020,Mehdi2021f,Bentley2013,Duan2004}. The improved results presented in this work can be ascribed to two key changes we made to the GPG optimization method as compared to Refs.~\cite{Gale2020,Mehdi2021f}. First, we add a penalty on the total number of SDKs such that the cost function for the first stage of the optimization takes the form:
\begin{align}
    J_1(\{z_k\}) = \varepsilon_{\rm av}\times c_1\left(\tanh(c_2[\mathcal{N}-(\mathcal{N}_{\rm max}+1)])+1 \right)
\end{align}
where $\varepsilon_{\rm av}$ is the state-averaged infidelity expression given by Eq.~\eqref{eq:Infidelity_Truncated}, $\mathcal{N}_{\rm max}$ is the cap on the number of SDKs, and $c_{1,2}$ are parameters that are chosen for rapid convergence of the stochastic local optimizations performed over the optimization variables $z_k$ of the first stage. For $\mathcal{N}\ll\mathcal{N}_{\rm max}$, the cost is simply the gate error $J\approx \varepsilon_{\rm av}$. We iteratively increase the value of $\mathcal{N}_{\rm max}$ until a high-fidelity solution ($\varepsilon_{\rm av}\leq 10^{-3}$) is found. Note that the second stage of the optimization does not penalise the number of SDKs, as changing the SDK timings $t_k$ does not change the sensitivity of 2Q gate scheme under our simple model of compounding pulse errors. \\
The second modification we make to the optimization approach is the use a larger number of `free' pulse groups in the initial search ($16-20$ as compared to $8-10$ in \cite{Gale2020,Mehdi2021}), with fewer pulses per group ($1-5$).  The increased dimensionality results in a sparsely sampled cost function for true global minima to be found, however we find that the increased dimensionality in the first stage of optimization provides more timing degrees of freedom which enables high-fidelity solutions to be identified that are compatible with a lower SDK bandwidth. The finite SDK bandwidth is supported by the addition of a $\tanh$ penalty on SDK timings in the second stage of the optimization to prevent pulse groups overlapping, i.e. the second stage uses a modified cost function:
\begin{align}
    J_2(\{t_k\}) = \varepsilon_{\rm av}+ \sum_{k\neq k'} c_1\left(\tanh(-c_2|t_k-t_{k'}|)+1 \right)
\end{align}
using $c_1 = 10$ and $c_2=10^{9}$. This latter aspect is key to finding high-fidelity fast gate solutions that are compatible with hyperfine qubits where SDKs are implemented using a few-nanosecond train of Raman pulses~\cite{Mizrahi2013,Mizrahi2013b,Johnson2015,Johnson2017}, which sets stricter bandwidth limits than for gates using single-pulse SDKs which can be implemented in picoseconds~\cite{Hussain2023,Heinrich2019,Guo2022d,Putnam2024a}.}

\subsubsection{Parameters and motional modes}

In the gate searches presented in the main text, we assumed the following wavelengths for the Raman lasers used to drive the SDKs on each ion: $532$nm for $^{133}$Ba and $^{138}$Ba~\cite{Putnam2024a}, $393$nm for $^{43}$Ca and $^{40}$Ca~\cite{Hussain2023a,Heinrich2019}, $369.5$nm for $^{171}$Yb~\cite{Guo2022d,Wong-Campos2017a,Mizrahi2013,Campbell2010a}, $408$nm for $^{88}$Sr, and $313$nm for $^{9}$Be. 

Each gate search requires computation of the secular mode frequencies, which can be solved analytically for a general mixed-species crystal of two ions~\cite{Wbbena2012}:
\begin{align}
    \omega_{\rm ip}^2 &= \omega_{0,1}^2\frac{1+\mu-\sqrt{1-\mu+\mu^2}}{\mu} \,, \\
    \omega_{\rm op}^2 &= \omega_{0,1}^2\frac{1+\mu+\sqrt{1-\mu+\mu^2}}{\mu} \,,
\end{align}
where $\mu=m^{(2)}/m^{(1)}$ is the mass ratio of the mixed-species pair, and $\omega_{0,1}$ is the secular trapping frequency of ion $1$ (taken to be $1$MHz in the calculations of the main text). The corresponding normal mode eigenvectors for the ip and op modes are given by $\mathbf{b}_{\rm ip}=(b_1,b_2)$ and $\mathbf{b}_{\rm op}=(b_1,-b_2)$, respectively, where
\begin{align}
    b_1^2 = \frac{1-\mu+\sqrt{1-\mu+\mu^2}}{2\sqrt{1-\mu+\mu^2}} \,,
\end{align}
and $b_2 = \sqrt{1-b_1^2}$.

\bibliographystyle{bibsty}
\bibliography{bib}

\pagebreak
\widetext
\renewcommand{\thefigure}{[S\arabic{figure}]} 
\renewcommand{\theequation}{S\arabic{equation}} 
\setcounter{equation}{0}
\setcounter{figure}{0}

\begin{center}

{\fontsize{11pt}{11pt}\selectfont\bfseries{Supplemental Materials for `Fast mixed-species quantum logic gates for trapped-ion quantum networks'}\par}

\end{center}
\vspace{0.1pt}

In this supplemental material we provide (1) a detailed derivation of the unitary operator and condition equations for the fast entangling gate mechanism (Eqs.~(1-2) of the main text), (2) a detailed derivation of the state-averaged infidelity (Eq.~(3) of the main text), and (3) additional error analysis of the fast gate mechanism under timing jitter and frequency shifts.

\section{(1) Gate unitary and phase gate condition equations}
Here we provided a detailed derivation of the unitary operator describing a heterogeneous fast gate protocol composed of $\mathcal{N}$ state-dependent kicks (SDKs) interspersed by free motional evolution, which generalises previous derivations for same-species systems~\cite{Garcia-Ripoll2003,Garcia-Ripoll2005}.

\subsection{State-dependent kicks}
An SDK can be realised using a pair of counter-propagating $\pi$-pulses. Each $\pi$-pulse performs a population invertion of the qubit state, alongside a photon recoil of $\pm \hbar k$: the unitary operator for a $\pi$-pulse incident on the $j$-th ion is described by the unitary operator:
\begin{align}
    U_\pi^{(j)}(k^{(j)}) = \sigma_- e^{i(k^{(j)}\hat{x}^{(j)}+\phi_L)} + h.c. = \sigma_x^{(j)} e^{i (k^{(j)} \hat{x}^{(j)}+\phi_L)\sigma_z^{(j)}} \,,
\end{align}
where $\phi_L$ is the laser phase. The counter-propagating $\pi$-pulse is described by the same unitary, with inverted wavevector $k^{(j)}\rightarrow -k^{(j)}$. We additionally assume the counter-propagating pulse to carry phase $-\phi_L$, such that the contribution of the laser phase cancels for each SDK~\footnote{This assumption is not required for the two-qubit entangling gate mechanism, as otherwise the phase of the SDK would be proportional to the relative phase between the counter-propagating pair, which can be characterised and thus eliminated with single-qubit rotations.}.
Assuming the time delay between the two pulses is much faster than the motion of the ions, an SDK incident on each of the two target ions can be written as:
\begin{align}
    U_{\rm SDK} \equiv U_\pi^{(2)}(-k^{(2)})U_\pi^{(1)}(-k^{(1)})U_\pi^{(2)}(k^{(2)})U_\pi^{(1)}(k^{(1)})  = e^{2i \left( k^{(1)}\hat{x}^{(1)}\sigma_z^{(1)} + k^{(2)}\hat{x}^{(2)}\sigma_z^{(2)}\right) } \,.
\end{align}
We then proceed by expanding the position operator of each ion in terms of quantized normal modes:
\begin{align}
\label{eq:x_MotionalModeExpansion}
\hat{x}^{(j)} = \sum_\alpha b_\alpha^{(j)} \sqrt{\frac{\hbar}{m^{(j)}\omega_\alpha}} \hat{X}_\alpha \,,
\end{align}
where $m^{(j)}$ is the mass of the $j$-th ion, $b_\alpha^{(j)}$ gives the coupling of the $j$-th ion to the $\alpha$-th normal mode, and we have defined the dimensionless position quadrature of the $\alpha$-th motional mode, $\hat{X}_\alpha\equiv (\hat{a}_\alpha+\hat{a}^\dag_\alpha)/\sqrt{2}$. 
Using Eq.~\eqref{eq:x_MotionalModeExpansion}, the SDK unitary can be written as a product of displacement operators on each motional mode, $\hat{D}_\alpha(\beta) = \exp{\beta \hat{a}^{\dag}_\alpha- \beta^* \hat{a}_\alpha }$, i.e. 
\begin{align}
\label{eq:SDKunitary_supp}
    U_{\rm SDK}(k^{(1)},k^{(2)}) &= \prod_{j=1,2}\prod_\alpha  \hat{D}_\alpha\left(2i b_\alpha^{(j)} \eta_\alpha^{(j)} \hat{\sigma}_z^{(j)}   \right) \,, 
\end{align}
where $\eta_\alpha^{(j)} = k^{(j)}\sqrt{\hbar/(2m^{(j)}\omega_\alpha)}$ is the mode- and ion-dependent Lamb-Dicke parameter.

It is then straightforward to describe the action of a single SDK on a general two-qubit state of the form:
\begin{align}
\label{eq:supp:twoqubitstate_initial}
    \ket{\psi_0} = \left( c_{\uparrow \uparrow}\ket{\uparrow\uparrow} +c_{\uparrow \downarrow}\ket{\uparrow \downarrow}+c_{\downarrow \uparrow}\ket{\downarrow \uparrow}+c_{\downarrow\downarrow}\ket{\downarrow\downarrow}\right)\otimes \ket{\psi_{\rm mot}} \,,
\end{align}
where we have assumed a pure motional state, $\ket{\psi_{\rm mot}}$, shared by all internal states. For the present study, where all operations are either displacements or rotations in phase, it is sufficient to consider the case where the motional state is a product of coherent states of each motional mode, i.e. $\ket{\psi_{\rm mot}}=\otimes_\alpha \ket{\beta_{\alpha,0} }$, where $\beta_{\alpha,0}$ is the initial coherent state amplitude of the $\alpha$-th motional mode. In this case, the kicked state $\ket{\psi_{\rm kicked}} = \hat{U}_{\rm SDK}\ket{\psi_0}$ becomes:
\begin{align}
    \ket{\psi_{\rm kicked}} = \bigotimes_\alpha&\bigg[\left( c_{\uparrow \uparrow} \ket{\uparrow\uparrow}\ket{\beta_{\alpha,0}+4i \bar{\eta}_\alpha^{+}}+c_{\downarrow\downarrow} \ket{\downarrow\downarrow}\ket{\beta_{\alpha,0}-4i \bar{\eta}_\alpha^{+}}  \right) \notag \\ \,  +&\left(c_{\uparrow \downarrow} \ket{\uparrow \downarrow}\ket{\beta_{\alpha,0}-4i \bar{\eta}_\alpha^{-}}+c_{\downarrow \uparrow}\ket{\downarrow \uparrow}\ket{\beta_{\alpha,0}+4i \bar{\eta}_\alpha^{-}} \right) \bigg] \,,
\end{align}
where we defined the effective Lamb-Dicke couplings to the centre-of-mass ($+$) and relative ($-$) coordinates:
\begin{subequations}
\label{eq:supp:Effective_LDCouplings}
    \begin{align}
        \bar{\eta}_\alpha^{+} &= \frac{1}{2}\left(b_\alpha^{(1)}\eta_\alpha^{(1)}+b_\alpha^{(2)}\eta_\alpha^{(2)} \right) \,, \\
        \bar{\eta}_\alpha^{-} &= \frac{1}{2}\left(b_\alpha^{(2)}\eta_\alpha^{(2)} -b_\alpha^{(1)}\eta_\alpha^{(1)} \right) \,.
\end{align}  
\end{subequations}

These parameters describe the effective coupling of the SDK to the same-spin two-qubit states (CM: $\{\ket{\uparrow\uparrow}$,$\ket{\downarrow\downarrow}\}$) and opposite spin-states (rel: $\{\ket{\uparrow\downarrow}$,$\ket{\downarrow\uparrow}\}$), respectively.

\subsection{Derivation of Eq.~(2) of the main text}
As derived in the main text, the unitary for the pulsed fast gate operation is given by a product of impulsive SDKs at times $t_m$ with wavevector $z_m k$ (where $z_m =\pm 1$ denotes the pulse direction), interspersed by free motional evolution:
\begin{align}
    \hat{U}_{\rm G} = \prod_{m=1}^{\mathcal{N}}  \hat{U}_{\rm free}(t_{m+1}-t_m)\hat{U}_{\rm SDK}(z_mk_1,z_mk_2) \,,
\end{align}
where $U_{\rm free}(t) = \prod_\alpha \exp{-i \omega_\alpha (\hat{a}_\alpha^\dag \hat{a}_\alpha+1/2) t}$ describes free harmonic evolution of each motional mode with (secular) mode frequency $\omega_\alpha$. The duration of the gate operation is set by the time between the first and last SDK; in ideal operation, the final free evolution period has no effect on the two-qubit entanglement, nor does free evolution prior to the first pulse. Therefore, we can equivalently express the gate unitary as:
\begin{align}
    \hat{U}_{\rm G} = \prod_{m=1}^{\mathcal{N}}  \hat{U}_{\rm SDK}(z_mk_1,z_mk_2)\hat{U}_{\rm free}(
    \delta t_m)\,,
\end{align}
where $\delta t_m = t_m - t_{m-1}$, and we can set $\delta t_1=0$ without loss of generality.

Noting that the SDK unitary, Eq.~\eqref{eq:SDKunitary_supp}, leaves the two-qubit state unchanged, the gate unitary must be diagonal in the two-qubit basis $\{\ket{\uparrow\uparrow},\ket{\uparrow\downarrow},\ket{\downarrow\uparrow},\ket{\downarrow\downarrow}\}$. Thus, it is sufficient to consider the action of $\hat{U}_{\rm G}$ on each two-qubit state independently.

Consider the action of the gate unitary on the same-spin state $\ket{\uparrow\uparrow}$. Assuming an the initial state given by Eq.~\eqref{eq:supp:twoqubitstate_initial}, the final motional state associated with this two-qubit state is given by a product of coherent states for each mode
\begin{align}
\label{eq:supp:UG_UpUpElement}
    \bra{\uparrow\uparrow}\hat{U}_{\rm G}\ket{\psi_0} = \prod_{m}\prod_\alpha \hat{D}_\alpha(4iz_m \bar{\eta}_\alpha^{+})e^{-i\omega_\alpha \hat{a}^\dag_\alpha\hat{a}_\alpha  \delta t_m} \left[\bigotimes_\alpha \ket{\beta_{\alpha,0}} \right] \,,
\end{align}
where $\delta t_m = t_{m+1}-t_{m}$. We may then consider each motional mode independently to derive a relation between $\beta_{\alpha}^{\uparrow\uparrow}(\tau_{\rm G})$ and $\beta_{\alpha,0}$, where $\tau_{\rm G}=\sum_m t_m$ is the gate operation time. We use several properties of displacement and rotation operations on coherent states -- $\ket{\beta} = \hat{D}(\beta)\ket{0}$ (where $\ket{0}$ is the vacuum, and we have dropped mode-dependent subscripts), $e^{-i\Theta \hat{a}^\dag\hat{a} }\ket{\beta}=\ket{\beta e^{-i\Theta}}$, and $\hat{D}(a)\hat{D}(b) = e^{i\textrm{Im}[ab^*]}\hat{D}(a+b)$ -- in the calculation that follows. First, we relate the motional state after $k$ iterations to the motional state after $k-1$ iterations, $\ket{\beta_{\alpha,k-1}}$, i.e.
\begin{align}
\label{eq:supp:SingleKick_afterIterations}
    \hat{D}_\alpha(4iz_k \bar{\eta}_\alpha^{+})e^{-i\omega_\alpha \hat{a}^\dag_\alpha\hat{a}_\alpha  \delta t_k} \ket{\beta_{\alpha,k-1}} = e^{i \Im{4iz_k\bar{\eta}_\alpha^{+}\beta_{\alpha,k-1}^*} }\ket{e^{-i\omega_\alpha \delta t_k}\beta_{\alpha,k-1}+4iz_k \bar{\eta}_\alpha^{+}} \,.
\end{align}
Then, to identify the final state after $\mathcal{N}$ SDKs, we need to figure out the complex amplitude of the final coherent state, as well as the accumulated phase due to displacements generated by non-commuting operators.

First, we find the final coherent state amplitude. From Eq.~\eqref{eq:supp:SingleKick_afterIterations}, we can identify the following recursion relation for the amplitude of each mode:
\begin{align}
    \beta_{\alpha,k} &= e^{-i\omega_\alpha \delta t_k}\beta_{\alpha,k-1}+4iz_k \bar{\eta}_\alpha^{+} \,, \\
    &=  4i\bar{\eta}_\alpha^{+}\sum_{n=1}^{k} z_k e^{-i\omega_\alpha\sum_{j=n}^k\delta t_j} + \beta_{\alpha,0}e^{-i\omega_\alpha t_k} \,,
\end{align}
where we have obtained the second line by expanding out the recursion relation in terms of the initial coherent state amplitude, $\beta_{\alpha,0}$. The second term describes free evolution of the motional mode if there were no SDKs -- as $\beta_{\alpha,0}=0$ can be taken without loss of generality, the first term thus describes the effect of the kicks. Then, noting $\sum_{j=n}^k \delta t_j = t_k - t_n$, we can identify the final coherent state amplitude:
\begin{align}
\label{eq:supp:RecursionRelation_11}
    \beta_{\alpha,\mathcal{N}} = \left(4i\bar{\eta}_\alpha^{+}\sum_{n=1}^\mathcal{N}z_n e^{i\omega_\alpha t_n}+\beta_{\alpha,0}\right)e^{-i\omega_\alpha \tau_{\rm G}} \,,
\end{align}
where $\tau_{\rm G}=\sum_n t_n$ is the gate operation time. Moving into the rotating frame of each of the motional modes, the final coherent state amplitude can then be written as:
\begin{align}
    \label{eq:supp:FinalCoherentState_11}
    \beta_{\alpha,\mathcal{N}}^{\rm (rot.)} = 4i\bar{\eta}_\alpha^{+}\sum_{n=1}^\mathcal{N}z_n e^{i\omega_\alpha t_n}+\beta_{\alpha,0} \,.
\end{align}
A similar calculation yields the final motional amplitudes for the other two-qubit states: 
\begin{align}
\label{eq:supp:MotionalAmplitudes_EachTwoQubitState}
    \beta_{\alpha,\mathcal{N}}^{\downarrow\downarrow} &= -\beta_{\alpha,\mathcal{N}}^{\uparrow\uparrow} \,, \\ 
    \beta_{\alpha}^{\downarrow\uparrow}&=4i\bar{\eta}_\alpha^{-}\sum_{n=1}^\mathcal{N}z_n e^{i\omega_\alpha t_n}+\beta_{\alpha,0} \,, \\
    \beta_{\alpha,\mathcal{N}}^{\uparrow\downarrow}&=-\beta_{\alpha,\mathcal{N}}^{\downarrow\uparrow}\,.
\end{align}
Henceforth we will adopt the shorthand notation $\beta_\alpha^+ \equiv \beta_{\alpha,\mathcal{N}}^{\uparrow\uparrow} = -\beta_{\alpha,\mathcal{N}}^{\downarrow\uparrow}$ and $\beta_\alpha^- \equiv \beta_{\alpha,\mathcal{N}}^{\downarrow\uparrow} = -\beta_{\alpha,\mathcal{N}}^{\uparrow\downarrow}$.

In order for the internal qubit states and motional degrees of freedom to decouple by the end of the gate operation, the final coherent state amplitude should be the same as if there were no SDKs and only free evolution for period $\tau_{\rm G}$~\cite{Garcia-Ripoll2003}. By inspection of the above equations, this implies the motional restoration condition:
\begin{align}
\label{eq:supp:motres}
    \sum_{n=1}^{\mathcal{N}} z_n e^{i\omega_\alpha t_n} =0 \,, \;\;\;\forall \alpha \,.
\end{align}

Next, we consider the phase accumulated by the $\ket{\uparrow\uparrow}$ two-qubit state during the gate operation. By inspection of Equation \eqref{eq:supp:UG_UpUpElement}, we can identify 
the two-qubit phase arising from the non-commutativity of displacement operators along non-commuting axes in phase space (see Eq.~\eqref{eq:supp:SingleKick_afterIterations}):
\begin{align}
    \Theta_\alpha^{\uparrow\uparrow} = 4\bar{\eta}_\alpha^{+}\sum_{m=1}^\mathcal{N}z_m \Im{i\beta_{\alpha,k-1}^*e^{i\omega_\alpha \delta t_m}} \,,
\end{align}
which is a mode-dependent quantity. Using the recursion relation for the coherent state amplitude, Eq.~\eqref{eq:supp:RecursionRelation_11}, the latter expression may be written as:
\begin{align}
     \Theta_\alpha^{\uparrow\uparrow} &= 4\bar{\eta}_\alpha^{+}\sum_{k=1}^{\mathcal{N}} z_k \bigg(4\bar{\eta}_\alpha^{+}\sum_{n=1}^{k-1} z_n\sin(\omega_\alpha[t_k-t_n]) + {\rm Re}[\beta_{\alpha,0}^*e^{i\omega_\alpha t_k}]         \bigg)\\
    &=16(\bar{\eta}_\alpha^{+})^2\sum_{m=2}^{\mathcal{N}}\sum_{n=1}^{m-1}z_nz_m\sin(\omega_\alpha[t_m-t_n])+8\bar{\eta}_\alpha^{+}\sum_{k=1}^{\mathcal{N}} z_k \;{\rm Re}[\beta_{\alpha,0}^*e^{i\omega_\alpha t_k}] \,.
\end{align}
Without loss of generality, we may choose the initial coherent state amplitude to be $\beta_{\alpha,0}=0$, in which case the latter term in the above expression vanishes. The total phase accumulated by the $\ket{\uparrow\uparrow}$ state is then given by summing over all motional modes:
\begin{align}
\label{eq:supp:TwoQubit_Phase_11}
    \Theta^{\uparrow\uparrow} = 16\sum_\alpha (\bar{\eta}_\alpha^{+})^2\sum_{m=2}^{\mathcal{N}}\sum_{n=1}^{m-1}z_nz_m\sin(\omega_\alpha[t_m-t_n]) \,.
\end{align}

An analogous calculation for the other two-qubit basis states yields the phases accumulated by each, 
\begin{subequations}
\label{eq:supp:TwoQubit_Phases}
    \begin{align}
        \Theta^{\downarrow\downarrow} &= \Theta^{\uparrow\uparrow} \,, \\
        \Theta^{\downarrow\uparrow} &= 16\sum_\alpha (\bar{\eta}_\alpha^{-})^2\sum_{m=2}^{\mathcal{N}}\sum_{n=1}^{m-1}z_nz_m\sin(\omega_\alpha[t_m-t_n]) \,, \\
        \Theta^{\uparrow\downarrow} &= \Theta^{\downarrow\uparrow} \,.
\end{align}
\end{subequations}
Then, assuming that the motional restoration condition, Eq.~\eqref{eq:supp:motres}, is satisfied, we may express the unitary for the gate in the two-qubit basis as the matrix:
\begin{align}
    \hat{U}_{\rm G} = \left(
\begin{array}{cccc}
 \exp (i \Theta^{\uparrow\uparrow}) & 0 & 0 & 0 \\
 0 & \exp (i \Theta^{\downarrow\uparrow}) & 0 & 0 \\
 0 & 0 & \exp (i \Theta^{\downarrow\uparrow}) & 0 \\
 0 & 0 & 0 & \exp (i \Theta^{\uparrow\uparrow})  \\
\end{array}
\right)= \exp (i \frac{\Theta^{\uparrow\uparrow}+\Theta^{\downarrow\uparrow}}{2})\exp{i\Theta\hat{\sigma}_z^{(1)}\otimes\hat{\sigma}_z^{(2)}} \,,
\end{align}
which is, up to a global phase, a $\sigma_z\otimes\sigma_z$ phase gate with relative phase $\Theta$ given by:
\begin{align}
    \Theta \equiv \frac{\Theta^{\uparrow\uparrow}-\Theta^{\downarrow\uparrow}}{2} &=8\sum_\alpha\left[(\bar{\eta}_\alpha^{+})^2-(\bar{\eta}_\alpha^{-})^2\right]\sum_{m=2}^{\mathcal{N}}\sum_{n=1}^{m-1}z_nz_m\sin(\omega_\alpha[t_m-t_n]) \,, \\
    &= 8\sum_\alpha b_\alpha^{(1)}b_\alpha^{(2)} \eta_\alpha^{(1)} \eta_\alpha^{(2)}\sum_{m=2}^{\mathcal{N}}\sum_{n=1}^{m-1}z_nz_m\sin(\omega_\alpha[t_m-t_n])\,.
\end{align}
We have substituted Eq.~\eqref{eq:supp:Effective_LDCouplings} to obtain the second line, and this expression is Eq.~(2) of the main text. 
It is straightforward to check that eliminating the ion dependence of the Lamb-Dicke parameter in  the above expression, i.e. setting $\eta_\alpha^{(1)}=\eta_\alpha^{(2)}\equiv \eta_\alpha$, recovers the same-species result~\cite{Garcia-Ripoll2003,Garcia-Ripoll2005,Bentley2012a,Bentley2013,Ratcliffe2018,Gale2020}:
\begin{align}
    \Theta = 8\sum_\alpha b_\alpha^{(1)}b_\alpha^{(2)} \eta_\alpha^2\sum_{m=2}^{\mathcal{N}}\sum_{n=1}^{m-1}z_nz_m\sin(\omega_\alpha[t_m-t_n])\,.
\end{align}

\section{(2) Fidelity derivation}
Ideally, fast gates implement a maximally-entangling $\sigma_z\otimes \sigma_z$ phase gate, described by the unitary operator
\begin{align}
	\hat{U}_{\rm id} =e^{i (\pi/4 )\sigma_z^{(1)}\sigma_z^{(2)}} = \left(
\begin{array}{cccc}
 e^{i\pi/4} & 0 & 0 & 0 \\
 0 &  e^{-i\pi/4} & 0 & 0 \\
 0 & 0 &  e^{-i\pi/4} & 0 \\
 0 & 0 & 0 &  e^{i\pi/4} \\
\end{array}
\right) \,,
\end{align}
up to a global phase. For a given two-qubit state, $\ket{\psi_0}$, the fidelity of a fast gate is given by 
\begin{align}
\label{eq:statedep_fidelity}
	\mathcal{F}_{\ket{\psi_0}} = {\rm Tr}_m\left[ \bra{\psi_0}\hat{U}_{\rm id}^\dag \hat{U}_{\rm gate} \ket{\psi_0}\bra{\psi_0}\otimes \hat{\rho}_m \hat{U}_{\rm gate}^\dag \hat{U}_{\rm id}\ket{\psi_0} \right]\,,
\end{align}
where the trace is over the motional degrees of freedom with initial state $\hat{\rho}_m$, and the unitary operator describing the fast gate operation is given by:
\begin{align}
	\hat{U}_{\rm gate} = \left(
\begin{array}{cccc}
 e^{i  \Theta^{\uparrow\uparrow}}\hat{D}_{\downarrow\downarrow} & 0 & 0 & 0 \\
 0 & e^{i \Theta^{\downarrow\uparrow}}\hat{D}_{\downarrow\uparrow} & 0 & 0 \\
 0 & 0 & e^{i \Theta^{\downarrow\uparrow}}\hat{D}_{\uparrow\downarrow} & 0 \\
 0 & 0 & 0 & e^{i  \Theta^{\uparrow\uparrow}}\hat{D}_{\uparrow\uparrow}  \\
\end{array}
\right) \,,
\end{align}
where
\begin{align}
    \Theta^{\uparrow\uparrow/\downarrow\uparrow}=16\sum_\alpha(\bar{\eta}_\alpha^{\pm})^2\sum_{m=2}^{\mathcal{N}}\sum_{n=1}^{m-1}z_nz_m\sin(\omega_\alpha[t_m-t_n])\,.
\end{align}

Then, moving into the rotating frame of each motional mode, we can describe the residual motion at the end of the gate by the displacement operators (noting that displacement operators for different modes commute):
\begin{subequations}
	\begin{align}
	\hat{D}_{\downarrow\downarrow}&=\left( \hat{D}_{\uparrow\uparrow} \right)^\dag = \prod_\alpha \hat{D}_\alpha( -\beta_{\alpha}^{+})\,, \\
	\hat{D}_{\downarrow\uparrow}&=\left( \hat{D}_{\uparrow\downarrow} \right)^\dag = \prod_\alpha \hat{D}_\alpha( -\beta_{\alpha}^{-})\,,
\end{align}
\end{subequations}
where we have defined the displacement amplitudes:
\begin{align}
	\beta_{\alpha}^{\pm} = 4i\bar{\eta}^{\pm}\sum_k z_k e^{i\omega_\alpha t_k} \,.
\end{align}

Then, using the cyclic nature of the trace, Eq.~\eqref{eq:statedep_fidelity} can be written as:
\begin{align}
	\mathcal{F}_{\ket{\psi_0}} =  {\rm Tr}_m\left[ \hat{A}^\dag \hat{A} \hat{\rho}_m \right] \equiv \langle \hat{A}^\dag \hat{A} \rangle_m\,,
\end{align}
where 
\begin{align}
	\hat{A} = \bra{\psi_0}\hat{U}_{\rm id}^\dag \hat{U}_{\rm gate} \ket{\psi_0} = e^{i(-\pi/4)}\left(P_{\downarrow\downarrow}\hat{D}_{\downarrow\downarrow}
+P_{\uparrow\uparrow}\hat{D}_{\uparrow\uparrow}\right)+e^{i(\Theta^{\downarrow\uparrow}+\pi/4)}\left(P_{\downarrow\uparrow}\hat{D}_{\downarrow\uparrow}
+P_{\uparrow\downarrow}\hat{D}_{\uparrow\downarrow}\right) \,,
\end{align}
and $P_{ij}$ is the probability of the two-qubit state $\ket{ij}$, satisfying the normalisation condition: 
\begin{align}
	\sum_{i,j=\uparrow,\downarrow} P_{ij} = 1\,.
\end{align}
Note that the fidelity is independent of the relative phases of the two-qubit states; this is an artefact assuming perfect $\pi$-pulses, such that each SDK exactly restores the internal state of the ions. The expression for the fidelity will then become a linear combination of pairs of displacement operators, 
 which can be simplified using the fact that motional operators on different modes commute to compute products of displacement operators, e.g.
\begin{align}
	\hat{D}_{\downarrow\downarrow}\hat{D}_{\downarrow\uparrow} &= \prod_\alpha \hat{D}_\alpha( -\beta_{\alpha}^{+})\hat{D}_\alpha( -\beta_{\alpha}^{-}) \,, \\
	&= \prod_\alpha \hat{D}_\alpha\left(-(\beta_{\alpha}^{+}+\beta_{\alpha}^{-})\right) \,, \\ 
	&=\prod_\alpha \hat{D}_\alpha\big(-4i\bar{\eta}^{(2)}\sum_k z_k e^{i\omega_\alpha t_k})\equiv \prod_\alpha \hat{D}_\alpha\left(-\beta_\alpha^{(2)}\right) \,.
\end{align}
 Here we have used the relation $\hat{D}(\alpha)\hat{D}(\beta) = e^{\alpha\beta^*-\alpha^*\beta}\hat{D}(\alpha+\beta)$ to obtain the second line -- noting $\beta_\alpha^{+}$ and $\beta_\alpha^{-}$ share the same complex phase -- and substituted $\bar{\eta}^{\pm}_\alpha = (\bar{\eta}^{(2)}\pm\bar{\eta}^{(1)})/2$ to obtain the final line. Similarly, other pairs of displacement operators can be simplified:
 \begin{align}
 	\hat{D}_{\uparrow\uparrow}\hat{D}_{\uparrow\downarrow} &= \left( \hat{D}_{\downarrow\downarrow}\hat{D}_{\downarrow\uparrow} \right)^\dag = \prod_\alpha \hat{D}_\alpha\left(\beta_\alpha^{(2)}\right) \,, \\
 	\hat{D}_{\uparrow\uparrow}\hat{D}_{\downarrow\uparrow} &= \prod_\alpha \hat{D}_\alpha\left(\beta_\alpha^{(1)}\right) \,, \\
 	\hat{D}_{\downarrow\downarrow}\hat{D}_{\uparrow\downarrow} &=\left( \hat{D}_{\uparrow\uparrow}\hat{D}_{\downarrow\uparrow} \right)^\dag =  \prod_\alpha \hat{D}_\alpha\left(-\beta_\alpha^{(1)}\right) \,.
 \end{align}
 
Using the above result, the fidelity expression becomes:
 \begin{align}
	\mathcal{F}_{\ket{\psi_0}} =& \underbrace{P_{\uparrow\uparrow}^2+P_{\uparrow\downarrow}^2+P_{\downarrow\uparrow}^2+P_{\downarrow\downarrow}^2}_{=1}  \\ \notag & 
	-i e^{-i\Theta^{\downarrow\uparrow}} \left\langle P_{\uparrow\uparrow} \left(\hat{D}_\alpha\left(\beta_\alpha^{(1)} \right) P_{\uparrow\downarrow}+\hat{D}_\alpha\left(\beta_\alpha^{(2)} \right) P_{\downarrow\uparrow}\right)+P_{\downarrow\downarrow} \left(\hat{D}_\alpha\left(-\beta_\alpha^{(2)} \right) P_{\uparrow\downarrow}+\hat{D}_\alpha\left( -\beta_\alpha^{(1)} \right) P_{\downarrow\uparrow}\right)\right\rangle_m\\ \notag & 
	+i e^{i\Theta^{\downarrow\uparrow}} \left\langle P_{\uparrow\uparrow} \left(\hat{D}_\alpha\left(-\beta_\alpha^{(1)} \right) P_{\uparrow\downarrow}+\hat{D}_\alpha\left(-\beta_\alpha^{(2)} \right) P_{\downarrow\uparrow}\right)+P_{\downarrow\downarrow} \left(\hat{D}_\alpha\left(\beta_\alpha^{(2)} \right) P_{\uparrow\downarrow}+\hat{D}_\alpha\left(\beta_\alpha^{(1)} \right) P_{\downarrow\uparrow}\right)\right\rangle_m \\ \notag & 
	+P_{\uparrow\uparrow} P_{\downarrow\downarrow} \left\langle \hat{D}_\alpha \left(-(\beta_\alpha^{(1)}+\beta_\alpha^{(2)}) \right)+\hat{D}_\alpha \left(\beta_\alpha^{(1)}+\beta_\alpha^{(2)} \right)\right\rangle_m\\ \notag & +P_{\uparrow\downarrow} P_{\downarrow\uparrow} \left\langle \hat{D}_\alpha \left(\beta_\alpha^{(1)}-\beta_\alpha^{(2)} \right)+\hat{D}_\alpha \left(\beta_\alpha^{(2)}-\beta_\alpha^{(1)} \right)\right\rangle_m\,.
\end{align}
 
 To evaluate the trace over the motional degrees of freedom, we assume an initial thermal product state at temperature $T$:
 \begin{align}
 	\hat{\rho}_m = \bigotimes_\alpha \left(1-e^{-\hbar\omega_\alpha/(k_B T)}\right)\sum_{n=0}^\infty \ket{n}\bra{n}e^{-n\hbar\omega_\alpha/(k_B T)} \,.
 \end{align}
From this state, we can compute the thermal expectation value of products of displacement operators on each mode, i.e.
 \begin{align}
\left \langle \prod_\alpha 	\hat{D}_\alpha(\beta_\alpha) \right \rangle_m = \exp\left\{-\sum_\alpha|\beta_\alpha|^2\left(\bar{n}_\alpha+\frac{1}{2}\right) \right\} \,,
 \end{align}
 with $\bar{n}_\alpha=\coth(\hbar\omega_\alpha/(k_B T))-1/2$ giving the average phonon occupation of the $\alpha$-th mode. Using this result, we can then express the state-dependent fidelity as:
 \begin{align}
 \label{eq:statedependent_fidelity_final}
 \notag	\mathcal{F}_{\ket{\psi_0}} =& 
 	2P_{\uparrow\uparrow}\bigg(P_{\uparrow\downarrow}\cos(2\Delta \Phi)e^{-\sum_\alpha (\bar{n}_\alpha+1/2)|\beta_\alpha^{(1)}|^2} 
 	+P_{\downarrow\uparrow}\cos(2\Delta \Phi)e^{-\sum_\alpha (\bar{n}_\alpha+1/2)|\beta_\alpha^{(2)}|^2} + P_{\downarrow\downarrow} e^{-\sum_\alpha (\bar{n}_\alpha+1/2)|\beta_\alpha^{(1)}+\beta_\alpha^{(2)}|^2} \bigg) \\ \notag & 
 	+ 2P_{\uparrow\downarrow}\left(P_{\downarrow\uparrow}e^{-\sum_\alpha (\bar{n}_\alpha+1/2)|\beta_\alpha^{(1)}-\beta_\alpha^{(2)}|^2} + P_{\downarrow\downarrow}\cos(2\Delta \Phi)e^{-\sum_\alpha (\bar{n}_\alpha+1/2)|\beta_\alpha^{(2)}|^2}\right) \\ &
    +2P_{\downarrow\uparrow}P_{\downarrow\downarrow}\cos( 2\Delta\Phi)e^{-\sum_\alpha (\bar{n}_\alpha+1/2)|\beta_\alpha^{(2)}|^2} + 1 \,.
 \end{align}
 Here we have defined the error in the relative phase:
 \begin{align}
 	\Delta \Phi \equiv \bigg|\frac{\Theta^{\uparrow\uparrow}-\Theta^{\downarrow\uparrow}}{2}\bigg|-\frac{\pi}{4} \,,
 \end{align}
 such that when $\Delta \Phi =0$, and the motional states are perfectly restored (i.e. $\beta_\alpha^{(i)} =0$), the ideal unitary $\hat{U}_{\rm id}$ is implemented up to global phase -- i.e. $\mathcal{F}=1$.
 
 We may then obtain the state-averaged fidelity for the two-qubit state by integrating over the surface of the 3-sphere spanned by the probabilities $P_{ij}$ -- noting that the relative phases between the two-qubit states do not contribute to the fidelity expression. This is achieved by the use of hyperspherical coordinates for the coefficients of each of the two-qubit states, which gives
 \begin{subequations}
 	 \begin{align}
 	\label{eq:hypersphericalcoordinates}
 	P_{\downarrow\downarrow} &= \cos^2(\theta_1) \,, \\
 	P_{\downarrow\uparrow} &= \sin^2(\theta_1)\cos^2(\theta_2) \,, \\
 	P_{\uparrow\downarrow} &= \sin^2(\theta_1)\sin^2(\theta_2)\cos^2(\theta_3) \,, \\
 	P_{\uparrow\uparrow} &= \sin^2(\theta_1)\sin^2(\theta_2)\sin^2(\theta_3) \,,
 \end{align}
 \end{subequations}
where $0<\Theta_{1,2}<\pi$, and $0\leq\theta_{3}<2\pi$. The state-averaged fidelity, $\mathcal{F}_{\rm av}$, is then given by integrating Eq.~\eqref{eq:statedependent_fidelity_final} over all angles:
\begin{align}
\label{eq:stateaveragefidelity_integral}
	\mathcal{F}_{\rm av} &= \frac{\int_0^\pi d\theta_1 \int_0^\pi d\theta_2 \int_0^{2\pi}d\theta_3 \sin^2(\theta_1)\sin(\theta_2)\; \mathcal{F}_{\ket{\psi_0}}[\theta_1,\theta_2,\theta_3]}{\int_0^\pi d\theta_1 \int_0^\pi d\theta_2 \int_0^{2\pi}d\theta_3 \sin^2(\theta_1)\sin(\theta_2) } \,,\\
	\label{eq:stateaveragefidelity_fullresult}
	&=\frac{1}{2} + \frac{\cos(2\Delta\Phi)}{6}\left(e^{-\sum_\alpha (\bar{n}_\alpha+1/2)|\beta_\alpha^{(1)}|^2}+e^{-\sum_\alpha (\bar{n}_\alpha+1/2)|\beta_\alpha^{(2)}|^2} \right) \\ \notag 
	&+\frac{1}{12}\left(e^{-\sum_\alpha (\bar{n}_\alpha+1/2)|\beta_\alpha^{(1)}-\beta_\alpha^{(2)}|^2}+e^{-\sum_\alpha (\bar{n}_\alpha+1/2)|\beta_\alpha^{(1)}-\beta_\alpha^{(2)}|^2}\right) \,.
	\end{align}
 
 As we are interested in small errors, we may expand the above expression to leading order in $\Delta\Phi$ and $\{|\beta_\alpha^{(i)}|\}$ to obtain a simple expression for the infidelity. We perform this expansion in terms of the centre-of-mass and relative coordinates for the motional state amplitudes, by substituting
\begin{align}
	\beta_\alpha^{(1),(2)} = \beta_\alpha^{-} \pm \beta_\alpha^+
\end{align}
into Eq.~\eqref{eq:stateaveragefidelity_fullresult}, and then taking the series expansion to give:
\begin{align}
\label{eq:infidelityfunction_modebasis_appendix}
	 	1-\mathcal{F}_{\rm av} &\approx \frac{2 |\Delta \Phi|^2}{3} + \frac{2}{3}\sum_\alpha\left(\bar{n}_\alpha+\frac{1}{2}\right)\left(|\beta_{\alpha}^{+}|^2+|\beta_{\alpha}^{-}|^2 \right) \,, 
\end{align}
which gives Eq.~(3) of the main text. As a consistency check, we can consider the special case of a same-species crystal for which $\eta_\alpha^+ = \eta_\alpha^{-} \equiv \eta_\alpha$:
\begin{align}
	 	1-\mathcal{F}_{\rm av} &\approx \frac{2 |\Delta \Phi|^2}{3} + \frac{2}{3}\sum_\alpha\left(\bar{n}_\alpha+\frac{1}{2}\right)\left(|\beta_{\alpha}^{+}|^2+|\beta_{\alpha}^{-}|^2 \right) \,, \\
	 	&= \frac{2 |\Delta \Phi|^2}{3} + \frac{4}{3}\sum_\alpha\left(\bar{n}_\alpha+\frac{1}{2}\right)\left[(b_\alpha^{(1)})^2+(b_\alpha^{(2)})^2 \right]\left|2\eta_\alpha\sum_k z_k e^{i\omega_\alpha t_k}\right|^2 \,,
\end{align}
which is precisely the result employed in Refs.~\cite{Bentley2013,Taylor2017,Ratcliffe2018,Gale2020,Ratcliffe2020,Mehdi2021f,Mehdi2020b:2D}.

\subsection{Justification of instantaneous SDK approximation}
\zm{The theoretical framework developed above and in previous works studying pulsed fast gates assumes the SDK duration can be taken to be instantaneous with respect to the secular ion dynamics. While a detailed quantitative study of the effect of finite SDK duration is beyond the scope of this work (and will be the focus of a future manuscript), here we briefly estimate the potential two-qubit gate error induced by erroneous phase accumulation of $\delta \phi\sim \omega_0 \tau_{\rm SDK}$ for each SDK in the gate, where $\tau_{\rm SDK}$ is the finite SDK duration and the motional frequencies $\omega_\alpha$ have been crudely approximated by the secular trapping frequency $\omega_0$ (taken to be $1$MHz in the main text). For a two-qubit fast gate with $\mathcal{N}$ SDKs, we then consider a worst-case-scenario where the erroneous phases compound to give an error in the two-qubit entangling phase of $\Delta\Phi\sim \mathcal{N}\omega_0\tau_{\rm SDK}$. Eq.~\eqref{eq:infidelityfunction_modebasis_appendix} demonstrates that for small errors, the (state-averaged) two-qubit gate error scales quadratically with phase errors, i.e. $\epsilon_{2\rm Q}\sim|\mathcal{N}\omega_0\tau_{\rm SDK}|^2$. In the main text we consider nanosecond SDK durations and MHz mode frequencies such that $\omega_0\tau_{\rm SDK}\sim 10^{-3}$; for gate solutions with $\mathcal{N}\sim 10$ SDKs, this suggests contributions to the two-qubit gate error at the $10^{-4}$ level.}

\section{(3) Additional error analysis}
\begin{figure}[t!]
\centering
\includegraphics[width=\linewidth]{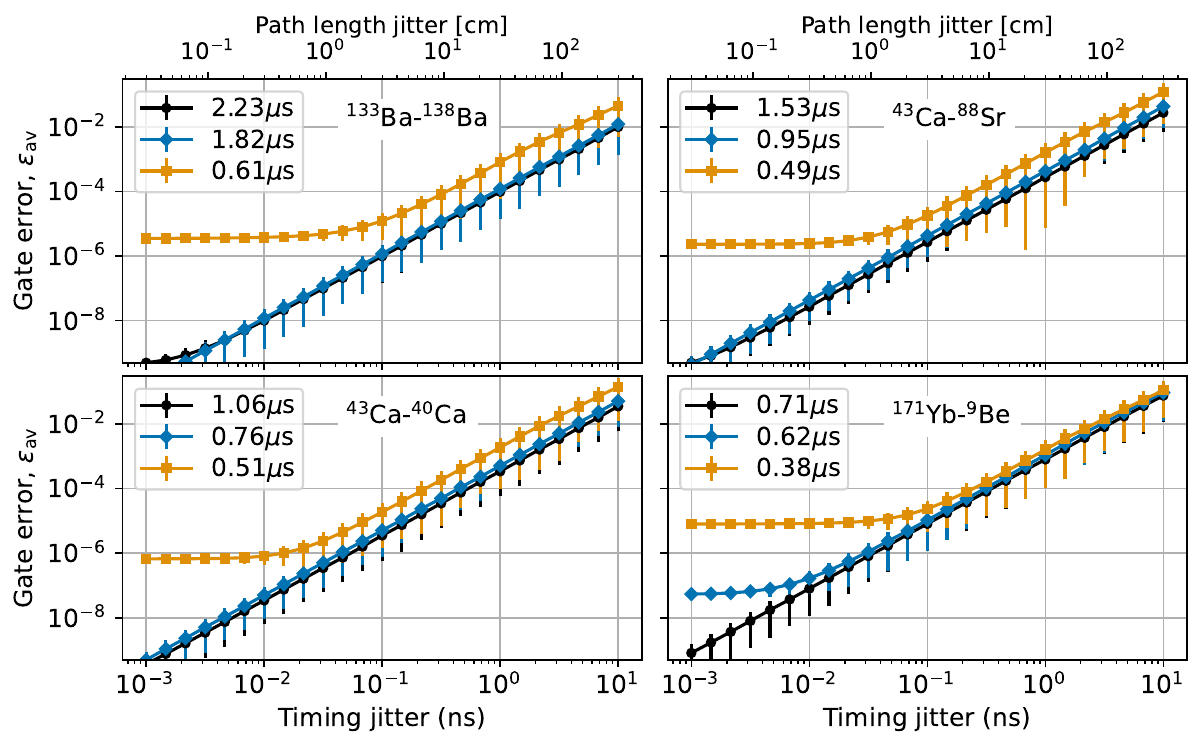}
    \caption{Sensitivity of MHz-speed gate solutions against timing jitter in the SDK pulse sequence, with the gate error averaged over $10^{4}$ noise realisations. Errorbars indicate the standard deviation in the ensemble average. }
    \label{fig:TimingJitter_Full}
\end{figure}

\begin{figure}[t!]
\centering
\includegraphics[width=\linewidth]{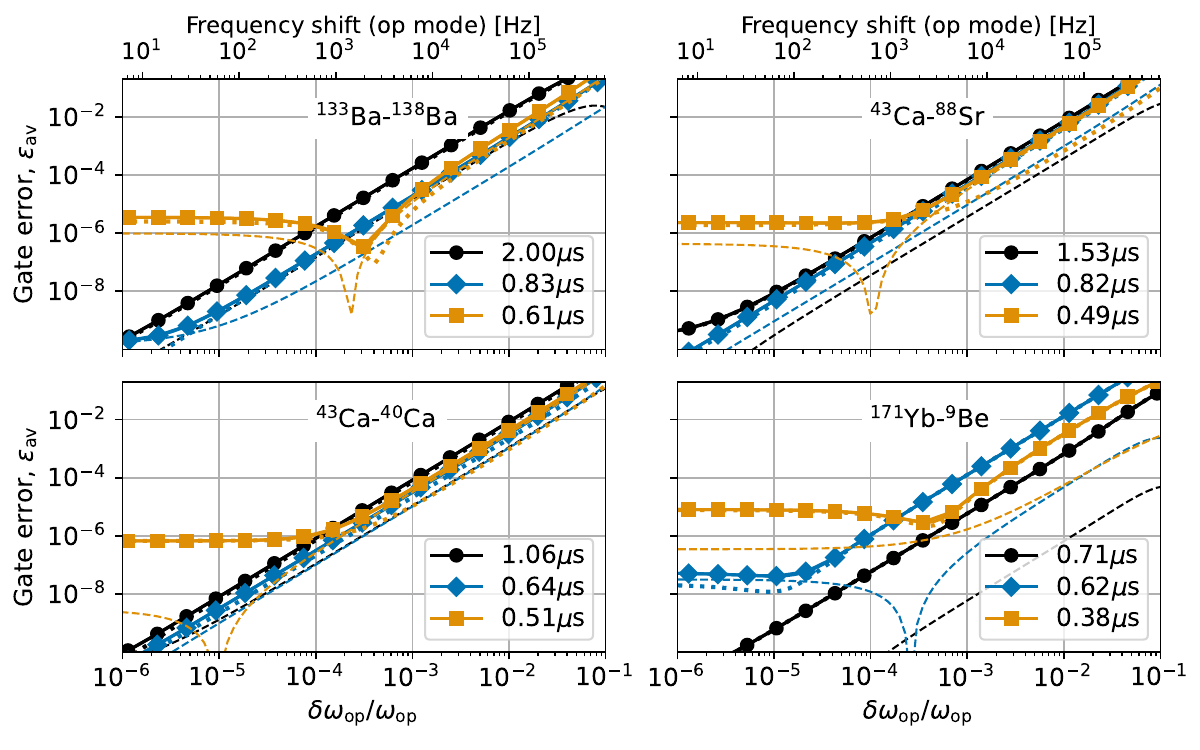}
    \caption{Effect of frequency drift on the out-of-phase (`op') motional mode, for a range of gate times. Dashed (dotted) lines indicates contribution of phase (motional) errors to the total gate infidelity given by Eq.~\eqref{eq:infidelityfunction_modebasis_appendix}. }\label{fig:OopFreqDrift_Full}
\end{figure}

\begin{figure}[t!]
\centering
\includegraphics[width=\linewidth]{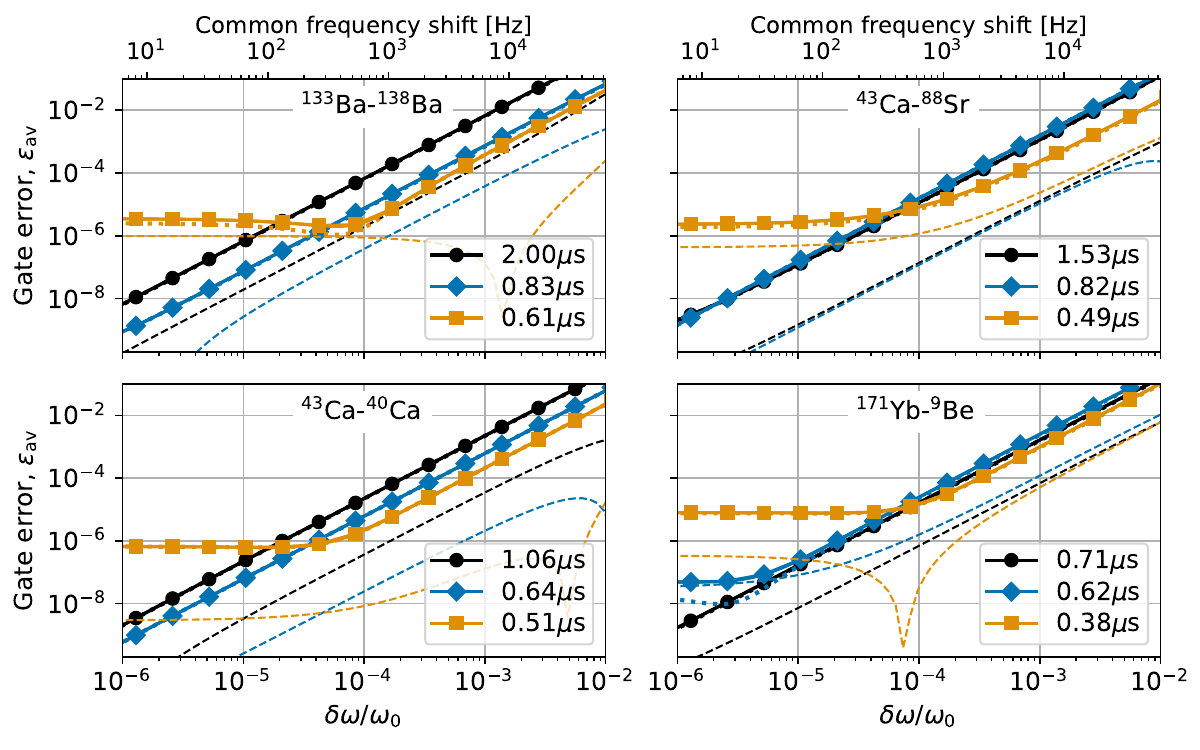}
    \caption{Effect of frequency drifts common to both motional modes, for a range of gate times. Dashed (dotted) lines indicates contribution of phase (motional) errors to the total gate infidelity given by Eq.~\eqref{eq:infidelityfunction_modebasis_appendix}.}
\label{fig:CommonFreqDrift_Full}
\end{figure}
In Figure~1(e) of the main text, we studied the effect of SDK timing jitter on select MHz-speed gates for the ion pairs $^{133}{\rm Ba}-^{138}$Ba and $^{43}{\rm Ca}-^{88}{\rm Sr}$. Figure~\ref{fig:TimingJitter_Full} provides a broader dataset that describes the effect of SDK timing jitter on gates between the dual-isotopes pair $^{43}{\rm Ca}-^{40}$Ca and the large mass-imbalance pair $^{171}{\rm Yb}-^{7}{\rm Be}$. The results are quantitatively and qualitatively similar to those discussed in the main text: subnanosecond timing control is sufficient to ensure gate errors of $\sim 10^{-4}$, for all ion pairs and gate speeds studied. \\
Figure~\ref{fig:OopFreqDrift_Full} similarly supplements Fig.~1(f) of the main text, wherein the impact of shifts in the frequency of the out-of-phase (op) mode is studied. The results are consistent across all four ion pairings studied: mode dependent shifts of $0.1\%$ ($1\%$) contribute gate errors of roughly $10^{-4}$ ($10^{-2}$) across all ion pairings and gate speeds studied. \\

Figure~\ref{fig:CommonFreqDrift_Full} presents results describing the degradation of gate fidelity under frequency shifts common to both modes, which is both qualitatively and quantitatively similar to the results in Fig.~\ref{fig:OopFreqDrift_Full} -- high-fidelity gate realisations require stabilisation of the trapping frequencies at the $0.1\%$ level, which translates to shifts of about $10$kHz for a $1$MHz secular trapping frequency.\\

\begin{figure}[t!]
\centering
\includegraphics[width=\linewidth]{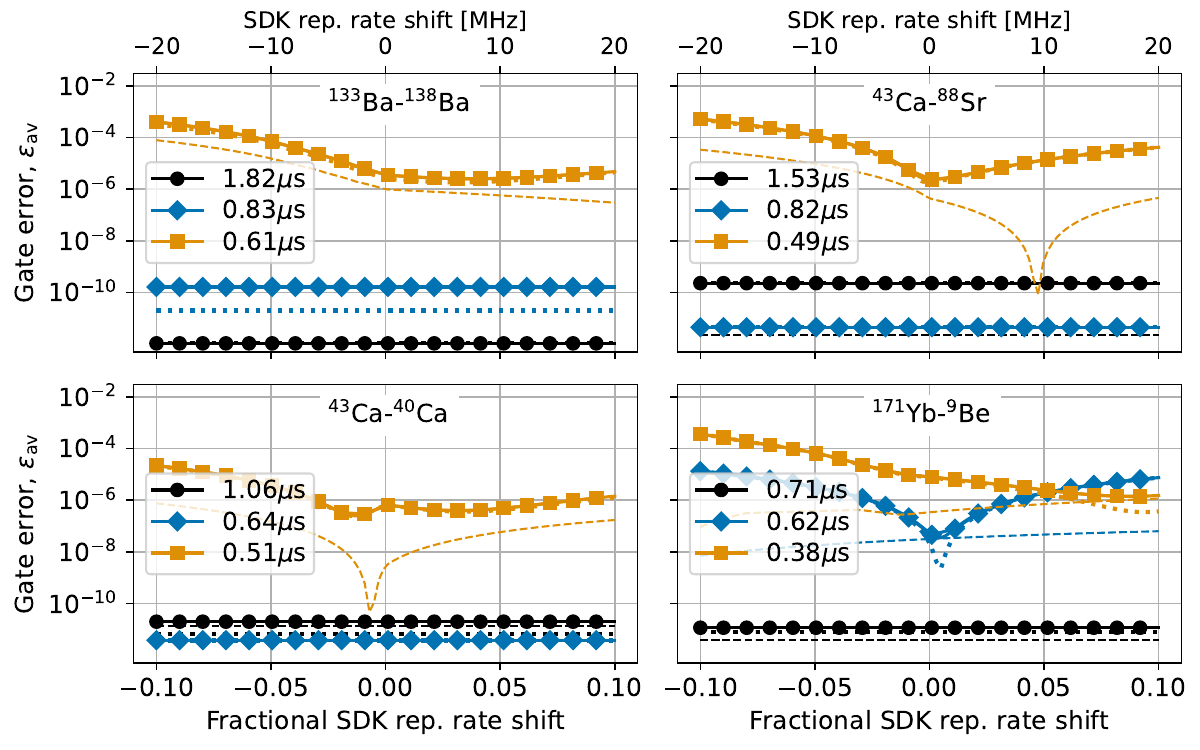}
    \caption{Effect of drifts or miscalibrations of the SDK repetition rate for gate solutions designed for a $200$MHz SDK bandwidth, for a range of gate times. Dashed (dotted) lines indicates contribution of phase (motional) errors to the total gate infidelity given by Eq.~\eqref{eq:infidelityfunction_modebasis_appendix}. }
\label{fig:SDKRepRateShifts}
\end{figure}

Lastly, we consider an error model to describe the gate error induced by drifts in the effective SDK repetition rate away from the $200$MHz SDK train bandwidth assumed in the main text. This error model is distinct to the effects of timing jitter as only the relative timings of SDKs separated by a single repetition period are modified -- this is treated as a systematic shift, rather than a random error. Fig.~\ref{fig:SDKRepRateShifts} demonstrates that gate solutions around $1\mu$s typically have SDKs separated by many repetition periods ($10-100$ns) and are thus entirely unaffected by drifts in the SDK repetition rate. We find that only gate solutions much faster than the trapping period, i.e. $\tau_{\rm G}\ll 1\mu$s, are affected by drifts in the SDK repetition rate, although large drifts of at least $10\%$ are required to contribute gate errors above $10^{-4}$. Interestingly, tuning the repetition rate by a few percent can improve the gate fidelity for some gate solutions.

\end{document}